# More period finding with adiabatic quantum computation

Richard H. Warren

*403 Bluebird Crossing, Glen Mills, PA 19342 USA*; richard.warren@villanova.edu

**Abstract** – We extend the work of Hen for the Bernstein-Vazirani problem and Simon's problem on an adiabatic quantum computer. Our results are based on the Ising objective function for quantum annealing. For each problem we determine its objective function, describe its Hamiltonian matrix, and show the Hamiltonian matrix for a small size problem. Following the pattern of Hen, we discuss factoring of integers, particularly Shor's factoring algorithm in an adiabatic quantum computing environment.

**Introduction.** – Significant developments are raising the expectations for adiabatic quantum computing. Probably the foremost development is the manufacture of adiabatic quantum computers (AQCs) that are capable of solving several types of minimization problems [1]. On another front, it has been shown theoretically that any algorithm for quantum circuit computing can be transferred in polynomial time to an AQC [2-3]. In addition, algorithms have been developed for adiabatic quantum computing and in some cases run on an AQC [4-8].

In this paper we continue the development of algorithms by bringing the work of Hen [4] closer to implementation on an AQC. Hen shows how an AQC can take the Bernstein-Vazirani problem [9] and Simon's problem [10] from an initial Hamiltonian $\hat{H}_d$ to a final Hamiltonian $\hat{H}_p$ using linear interpolation for a parameter varying smoothly with time. If this process is done slowly, then the system stays close to the ground state and the result is a state close to $\hat{H}_p$ that encodes the solution with high probably. This approach encapsulates the underlying theory for adiabatic quantum computing.

Our work is based on the Ising objective function for quantum annealing (1). We show how to write the Bernstein-Vazirani problem and Simon's problem in the form of (1). The coefficients $J_{ij}$ and $h_i$ in (1) become the entries in a Hamiltonian matrix $\hat{H}_d$ that is entered in an AQC. Then the AQC assigns 0, 1 to the $s_i$ in (1) so that a minimum is obtained. The output $\hat{H}_p$ encodes these values of $s_i$.

Hen [4] has a valuable discussion about Shor's factoring algorithm [11] and the work needed to implement it on an AQC. Since this is a major, worthwhile effort, we continue the discussion. Also, we have a negative comment about quantum factoring by multiplication.

**Adiabatic quantum computing.** – Adiabatic quantum computing is based on the adiabatic theorem of quantum mechanics [12] which says that under a slowly changing state an initial Hamiltonian will change to a final Hamiltonian representing its optimal state.

Quantum annealing is a type of adiabatic quantum computing where the qubits and their connections achieve an optimal state of low energy when super cooled. The Ising objective function for this optimal state is

$$\min\left(\sum_{i<j} s_i J_{ij} s_j + \sum_i h_i s_i\right) \quad (1)$$

where $i$ and $j$ are qubits, $s_i$ is the state of qubit $i$ (either 0 or 1), $h_i$ is the energy bias for qubit $i$, and $J_{ij}$ is the coupling energy between qubits $i$ and $j$. Quantum annealing can be thought of as evolution from an initial state of the $s_i$ to their final state according to weights $h_i$ and $J_{ij}$, and minimizes energy.

The original Ising model uses $\sigma_i \in \{-1, 1\}$ in place of $s_i$, where $\sigma_i$ represents spin. The relationship $\sigma_i = 2s_i - 1$ equates the models.

**The Bernstein-Vazirani problem.** – We will extend the work in [4] to an objective function and a Hamiltonian matrix that is ready for input to an AQC, such as manufactured by D-Wave Systems [13-14]. According to [4], in the Bernstein-Vaziran problem, one is given a black box that evaluates the function

$$f(w) = (\sum_{k=0}^{n-1} w_k a_k) \mod 2 \quad (2)$$

where $w_k$ and $a_k$ ($k = 0, \ldots, n-1$) are the bits of the two integers $w$ and $a$, respectively, and the function $f(\cdot)$ takes $w$ into the modulo-2 sum of the products of the corresponding bits of $w$ and $a$. The task is to find $a$ with as few queries of $f(\cdot)$ as possible.

First we note that $f(2^k w_k) = w_k a_k$, i.e., the bits of the product $wa$ can be calculated. The horizontal lines in Table 1 show that if a value for $a_k$ is given, then $w_k - 2w_k a_k$ attains its minimum if and only if $w_k = a_k$. So we will let the quantum annealing process minimize $w_k - 2w_k a_k$ by assigning a 0, 1 value to $w_k$. Thus, $w_k$ has the role of $s_i$ in (1).

Table 1: Evaluating $w_k - 2w_k a_k$

|  | if $w_k = 1$ | if $w_k = 0$ |
|---|---|---|
| if $a_k = 1$ | $w_k - 2w_k a_k = -1$ | $w_k - 2w_k a_k = 0$ |
| if $a_k = 0$ | $w_k - 2w_k a_k = 1$ | $w_k - 2w_k a_k = 0$ |

Therefore, $\min \sum_{k=0}^{n-1}(w_k - 2w_k a_k) \quad (3)$

is an objective function for the Bernstein-Vazirani problem that has the form of (1). The 0, 1 variables $w_k$ are assigned by the quantum annealing process and are the output. We observe that $O(1)$ queries are made, as in [4,9]. Next we show a Hamiltonian matrix for the objective function (3) when $n = 4$.

|  | $w_0$ | $w_1$ | $w_2$ | $w_3$ |
|---|---|---|---|---|
| $w_0$ | $1 - 2a_0$ | 0 | 0 | 0 |
| $w_1$ | 0 | $1 - 2a_1$ | 0 | 0 |
| $w_2$ | 0 | 0 | $1 - 2a_2$ | 0 |
| $w_3$ | 0 | 0 | 0 | $1 - 2a_3$ |

Thus, the input is an $n \times n$ matrix with $1 - 2a_k$ on the diagonal and 0 off the diagonal. Since the current adiabatic quantum computer manufactured by D-Wave Systems has 512 qubits and since the above Hamiltonian matrix is diagonal, a 512 bit integer problem can be solved if all qubits are operational.

Modulo-2 in (2) has no role in the best classical algorithm described in [4] and in our quantum adiabatic algorithm. Our algorithm uses quantum parallelism where $n$ subsystems can evolve without affecting each other. Thus, the runtime for the evolution of each subsystem is not dependent on the size of the system. As a result, runtime does not scale with system size.

**Simon's problem.** – We will replace the adiabatic quantum solution in [4] with an entirely new approach, and conclude with an objective function and a Hamiltonian matrix that is ready for input to an AQC manufactured by D-Wave Systems [13-14].

There are several variations of Simon's problem. In the [4] version we are told there is an $n$-bit positive integer $a$ such that for any two $n$-bit inputs $w \neq y$, a black-box function $g(\cdot) : \{0,1\}^n \to \{0,1\}^{n-1}$ outputs the $(n-1)$-bit integers $g(w) = g(y)$ if and only if $w \oplus y = a$. The symbol $\oplus$ denotes the bitwise operation xor. The problem is to find $a$ with as few queries of $g(\cdot)$ as possible.

First we note that the size of the domain of $g(\cdot)$ is $2^n$ and the size of its range is $2^{n-1}$. Therefore, there are at least $2^n - 2^{n-1}$ $n$-bit integers $w \neq y$ such that $g(w) = g(y)$. We will use the quantum annealing process to find such a pair and then determine $a$ from the truth table for $w \oplus y$ bitwise.

Next we note that Simon's problem does not tell us about the block-box output when $g(w) \neq g(y)$. So we assume that there is a signal of some sort to indicate this. We transcribe the signal to $g(w) - g(y) = 1$, and will use $g(w) - g(y)$ as the objective function to be minimized by the quantum annealing process. Thus, we will search for a pair of $n$-bit integers $w, y$ such that $g(w) - g(y) = 0$.

Lastly, a constraint is needed in our search to ensure $w \neq y$. Let $w_i$ and $y_i$ ($i = 1, \ldots, n$) be the bits of the integers $w$ and $y$, respectively. Fix $j \in \{1, \ldots, n\}$. We designate $w_j - y_j = 1$ for this constraint. Then we expand $(w_j - y_j - 1)^2 = 0$ and simplify. The result is a constraint function $-w_j + 3y_j - 2w_j y_j$, which attains its minimum when $w_j > y_j$.

We add the objective function and constraint function to obtain

$$\min(g(w) - g(y) - w_j + 3y_j - 2w_j y_j) \quad (4)$$

as the function to be solved by an AQC. The variables are $g(w), g(y), w_i$ and $y_i$ for $i \in \{1, \ldots, n\}$.

The average number of calls to an AQC to find a solution for (4) is $n/2$. This compares favorably to the $n - 1$ calls in [4] for Simon's problem.

The coefficients in (4) are the entries in a Hamiltonian matrix that is the input to an AQC. We show this matrix for $n = 3$ and $j = 1$.

|      | $w_1$ | $w_2$ | $w_3$ | $y_1$ | $y_2$ | $y_3$ | $g(w)$ | $g(y)$ |
|------|-------|-------|-------|-------|-------|-------|--------|--------|
| $w_1$ | -1   |       |       | -2    |       |       |        |        |
| $w_2$ |      | 0     |       |       |       |       |        |        |
| $w_3$ |      |       | 0     |       |       |       |        |        |
| $y_1$ | -2   |       |       | 3     |       |       |        |        |
| $y_2$ |      |       |       |       | 0     |       |        |        |
| $y_3$ |      |       |       |       |       | 0     |        |        |
| $g(w)$ |     |       |       |       |       |       | 1      |        |
| $g(y)$ |     |       |       |       |       |       |        | -1     |

The final step is to find $a_i$ by means of $a_i = w_i \oplus y_i$ for $i \in \{1, \ldots, n\}$. This can be done on a classical computer. The xor operation $\oplus$ (also known as exclusive or) is fully developed in [7] for adiabatic quantum computing.

A current AQC has 512 qubits, which translates to solving Simon's problem for integers with up to 255 bits, when all qubits are operational.

**Integer Factorization.** – We continue the discussion in [4] about quantum-adiabatic factoring of integers in polynomial time. As Hen points out, the work in [2-3] implies that Shor's factoring algorithm which is polynomial for a quantum circuit model computer has a polynomial time counterpart on an AQC.

Shor's factoring algorithm employs quantum speedup for period finding. To do this, Shor uses a Fourier transform to change the domain to roots of unity. After finding the indicator for period in the new domain, Shor uses an inverse Fourier transform to return to the original domain. It is known that a Hadamard gate can replace the Fourier transform, but apparently not the inverse. As indicated in [4], the Fourier transform and period finding via roots of unity are not in place for quantum adiabatic computing.

Furthermore, implementation of Shor's factoring algorithm on quantum circuit model apparatus [15-17] has used a sequence of steps where the output from one step is part of an input to the next step. This was proposed for adiabatic quantum computing in [7], but has not been demonstrated on an AQC, as far as we know.

The goal is to factor large integers [18] in polynomial time on an AQC. This will require larger AQCs. There are physical barriers to overcome in order to produce an AQC with more qubits and with a larger percent of connections between qubits [14 page 7].

The recent approach to quantum factoring by multiplication [19-20] has a significant limitation. Apparently, the complexity of classical multiplication is not known, but is widely believed to be exponential. The speed-up by quantum multiplication is an open question [19 page 110]. Thus, we do not recommend this approach.

**Summary and conclusions.** – We have provided the details to input the Bernstein-Vazirani problem and Simon's problem to an AQC. Our results for the first problem extend Hen's work [4] and have the same complexity. Our algorithm for Simon's problem is a new approach that exploits the minimization feature of the Ising function (1). Its complexity improves Hen's [4].

Algorithm construction, such as above, is one of the three legs supporting adiabatic quantum computing. Thus, the implementation of these algorithms on an AQC is an important next step and is expected to be a further demonstration of the potential of adiabatic quantum computing.

We agree with Hen's argument [4] that Shor's factoring algorithm [11] on an AQC is an important and significant goal for the future. So we have commented about the steps needed to implement it.

**Acknowledge.** – I express my deep appreciation to my Lord and Savior, Jesus Christ, who gave me the desire, ability, resources and strength to research this topic and write this paper.